\documentclass[preprint,prc,showpacs,preprintnumbers,
               superscriptaddress,amsmath,amssymb,floatfix]{revtex4}

\usepackage{graphicx}
\usepackage{dcolumn}
\usepackage{bm}
\usepackage{longtable}

\newcommand{\beq}{\begin{equation}}
\newcommand{\eeq}{\end{equation}}
\newcommand{\beqn}{\begin{eqnarray}}
\newcommand{\eeqn}{\end{eqnarray}}
\newcommand{\btab}{\begin{tabular}}
\newcommand{\etab}{\end{tabular}}
\newcommand{\brho}{\mbox{\boldmath$\rho$}}
\newcommand{\btau}{\mbox{\boldmath$\tau$}}

\begin{document}
\title{Density Dependencies of Interaction Strengths and their
Influences on Nuclear Matter and Neutron Stars in the Relativistic
Mean Field Theory}

\author{S. F. Ban}
\affiliation{School of Physics, Peking University, Beijing 100871,
China}

\author{J. Li}
\affiliation{School of Physics, Peking University, Beijing 100871,
China}
\affiliation{College of Physics and Technology, Wuhan
             University,~~Wuhan~~430072, China}

\author{S. Q. Zhang}
\affiliation{School of Physics, Peking University, Beijing 100871,
China}

\author{H. Y. Jia}
\affiliation{School of Physics, Peking University, Beijing 100871,
China}
\affiliation{College of Science, Southwest Jiaotong
              University,~~Chengdou~~610031, China}

\author{J. P. Sang}
\affiliation{College of Physics and Technology, Wuhan
             University,~~Wuhan~~430072, China}

\author{J. Meng}
\thanks{e-mail: mengj@pku.edu.cn}
\affiliation{School of Physics, Peking University, Beijing 100871,
China}
\affiliation{Institute of Theoretical Physics, Chinese
Academy of Science, Beijing 100080, China} \affiliation{Center of
Theoretical Nuclear Physics, National
             Laboratory of Heavy Ion Accelerator, Lanzhou 730000, China}

\date{\today}

\begin{abstract}
The density dependencies of various effective interaction
strengths in the relativistic mean field are studied and carefully
compared for nuclear matter and neutron stars. The influences of
different density dependencies are presented and discussed on mean
field potentials, saturation properties for nuclear matter,
equations of state, maximum masses and corresponding radii for
neutron stars. Though the interaction strengths and the potentials
given by various interactions are quite different in nuclear
matter, the differences of saturation properties are subtle,
except for NL2 and TM2, which are mainly used for light nuclei,
while the properties by various interactions for pure neutron
matter are quite different. To get an equation of state for
neutron matter without any ambiguity, it is necessary to constrain
the effective interactions either by microscopic many-body
calculations for the neutron matter data or the data of nuclei
with extreme isospin. For neutron stars, the interaction with
large interaction strengths give strong potentials and large
Oppenheimer-Volkoff (OV) mass limits. The density-dependent
interactions DD-ME1 and TW-99 favor a large neutron population due
to their weak $\rho$-meson field at high densities. The OV mass
limits calculated from different equations of state are 2.02
$\sim$ 2.81$ M_\odot$, and the corresponding radii are 10.78
$\sim$ 13.27 km. After the inclusion of the hyperons, the
corresponding values become 1.52 $\sim$ 2.06 $M_\odot$ and 10.24
$\sim$ 11.38 km.
\end{abstract}


\maketitle

\section{Introduction}

A widely used and successful approach for nuclear matter and
finite nuclei is the mean field theory employing effective
interactions. The mean field theory includes non-relativistic mean
field theory with effective nucleon-nucleon interactions such as
Skyrme or Gogny, and the relativistic mean field (RMF) theory.  In
a certain sense, the RMF theory is more fundamental as it starts
from a phenomenological hadronic field theory with strongly
interacting baryons and mesons as degrees of freedom
\cite{wale74}. It has been used not only for describing the
properties of nuclei near the valley of stability successfully
\cite{ring96}, but also for predicting the properties of exotic
nuclei with large neutron or proton excess \cite{meng98a,meng98b}.
In the mean field theory, the effective interactions are adjusted
to various properties of nuclear matter and finite nuclei. In
recent years, a number of effective interactions of meson-baryon
couplings based on the RMF theory have been developed, including
nonlinear self-couplings for the $\sigma$-meson and/or
$\omega$-meson, such as NL1, NL2 \cite{joon86}, NL3 \cite{lala97},
NLSH \cite{shar93}, TM1 and TM2 \cite{suga94}. However, these
nonlinear interactions have problems of stability at high
densities, as well as the question of their physical foundation
\cite{joon86}. A more natural alternative is to introduce the
density dependence in the couplings \cite{type99}. Based on the
Dirac-Brueckner calculations, Typel and Wolter proposed the
density-dependent effective interaction TW-99 and expected that
the model could be reasonably extrapolated to extreme conditions
of isospin and/or density \cite{type99}. Along this line,
Nik\v{s}i\'{c} et al. developed another effective interaction
DD-ME1 \cite{niks02}. In this paper, we will analyze the density
dependencies of various effective interactions including both the
nonlinear and density-dependent versions in the RMF theory and
investigate their influences on properties of nuclear matter and
neutron stars.

The existence of neutron stars was predicted following the
discovery of neutron. In 1934, Baade and Zwicky suggested that
neutron stars could be formed in ``supernovae" \cite{baad34}. The
radio pulsars discovered by Bell and Hewith in 1967 \cite{bell68}
were identified as rotating neutron stars by Pacini \cite{paci67}
and Gold \cite{gold69}. The first theoretical calculation of
neutron stars was performed by Oppenheimer and Volkoff
\cite{oppe39}, and independently by Tolman \cite{tolm39}. In their
calculation the neutron stars were assumed as gravitationally
bound states of neutron Fermi gas. Recent progress on the study of
neutron stars can be found in Refs. \cite{glen97,heis00} and
references therein.

The physics of neutron stars has offered an intriguing interplay
between nuclear processes and astrophysical observation, and has
become a hot topic in nuclear physics and astrophysics. The
neutron stars exhibit conditions far from those encountered on the
earth. The neutron star models including various so-called
realistic equations of state have resulted in the following
general picture for the interior of neutron star. The surface of
neutron star is a solid crust of thickness about 1 km, which is
mainly made up of nuclei and free electrons. Inside the crust,
charge-neutral neutron star mainly consists of neutrons together
with a small concentration of protons and electrons in equal
number. Proton and electron densities increase with total baryon
density, and the $\mu^-$, $\pi$, $K$ mesons and other baryons (
e.g., hyperons ) as well as a phase transition from baryon degrees
of freedom to quark matter will appear \cite{heis00}.

The equation of state (EOS) is essential to understand the
structure and properties of neutron stars. The EOS determines
properties such as the mass range, the mass-radius relationship,
the crust thickness, the cooling rate and even the energy released
in a supernova explosion. Usually, the EOS is obtained by
extrapolating the theory, which is developed mainly for normal
nuclear matter, to nuclear matter with extreme high isospin and
high densities. Unfortunately, such extrapolation is always
model-dependent. The RMF theory has proved to be very successful
in describing the properties of nuclear matter and finite nuclei
\cite{wale74,ring96}, rotation nuclei \cite{ring96}, nuclei far
from $\beta$ stability \cite{meng98a,meng98b} and magnetic
rotation \cite{mado00}, etc. Based on the RMF theory, there are
also many studies on neutron stars and strange nuclear matter with
effective interactions including nonlinear self-couplings for
scalar and vector mesons [17, 20-30].

In this paper, the density dependencies of various effective
interaction strengths in the RMF theory are studied and compared
for nuclear matter and neutron stars. The corresponding influences
of different density dependencies for effective interactions are
presented and discussed on mean field potentials, saturation
properties for nuclear matter, EOS, maximum masses and
corresponding radii for neutron stars. In Section II, a brief
description of the RMF theory in nuclear matter and neutron stars
is presented. The results and discussions are given in the
following section. In the last section, we give a brief summary.
\newpage
\section{A Sketch of the RMF theory in Nuclear matter and neutron stars}
The details of RMF theory can be found in a number of reviews
\cite{wale74,ring96,glen97}. The RMF theory starts from an
effective Lagrangian density with baryons, mesons ($\sigma$,
$\omega$ and $\rho$) and photons as degrees of freedom
($\hbar=c=1$):
 \beqn
  \label{lb}
   {\cal L}
   &=&\sum_B\bar{\psi}_B\left[i{\gamma^\mu}{\partial_\mu}-m_B-g_{\sigma B}\sigma
     -g_{\omega B}\gamma^\mu\omega_\mu - g_{\rho B}\gamma^\mu \btau_B\cdot\brho_\mu
     -e\gamma^\mu A_\mu\frac{1-\tau_{3B}}{2}\right]\psi_B\nonumber\\
   &&+\frac{1}{2}\partial_\mu\sigma\partial^\mu\sigma-\frac{1}{2}m_\sigma^2\sigma^2-U(\sigma)
     -\frac{1}{4}\omega_{\mu\nu}\omega^{\mu\nu}
     +\frac{1}{2}m_\omega^2\omega_\mu\omega^\mu+U(\omega)\nonumber\\
   &&-\frac{1}{4}\brho_{\mu\nu}\cdot\brho^{\mu\nu}
     +\frac{1}{2}m_\rho^2\brho_\mu\cdot\brho^\mu
     -\frac{1}{4}A_{\mu\nu}A^{\mu\nu},
 \eeqn
where the Dirac spinor $\psi_B$ denotes the baryon $B$ with mass
$m_B$ and isospin $\tau_B$. The sum on $B$ is over protons,
neutrons and hyperons ($\Lambda, \Sigma^\pm, \Sigma^0, \Xi^-,
\Xi^0$, et. al.) in this paper. The scalar sigma ($\sigma$) and
vector omega ($\omega$) offer medium-range attractive and
short-range repulsive interactions respectively, and the isospin
vector rho ($\brho$) provides the necessary isospin asymmetry.
Their masses are denoted by $m_\sigma,m_\omega$ and $m_\rho$. The
corresponding coupling meson-baryon constants are $g_{\sigma
B},g_{\omega B}$ and $g_{\rho B}$ respectively. $\btau_B$ is the
isospin of baryon $B$ and $\tau_{3B}$ is its 3-component. The
nonlinear self-couplings for $\sigma$ and $\omega$ mesons are
respectively:
 \beqn
   U(\sigma)=\frac{1}{3}g_2\sigma^3+\frac{1}{4}g_3\sigma^4,~~~~~
   U(\omega)=\frac{1}{4}c_3(\omega^\mu\omega_\mu)^2,
 \eeqn
with the self-coupling constants $g_2,g_3$, and $c_3$. The field
tensors $\omega_{\mu\nu}$, $\brho_{\mu\nu}$, and $A_{\mu\nu}$ are
for $\omega$-meson, $\rho$-meson, and photon respectively.

The parameterization of the interaction in the RMF theory is
obtained by fitting the properties of nuclear matter and some
finite nuclei. Instead of the self-coupling of the meson fields,
Typel and Wolter proposed the density dependencies of the
couplings in the RMF theory \cite{type99}, i.e., the coupling
constant $g_{\sigma(\omega)B}$ of the $\sigma$ ($\omega$) meson is
replaced by:
 \beqn
  \label{grho0}
    g_{\sigma(\omega)B}(\rho)=g_{\sigma(\omega)B}(\rho_0)f_{\sigma(\omega)}(x),
 \eeqn
where
 \beqn
 \label{fx}
  f_{\sigma(\omega)}(x)=a_{\sigma(\omega)}
    \frac{1+b_{\sigma(\omega)}(x+d_{\sigma(\omega)})^2}{1+c_{\sigma(\omega)}(x+d_{\sigma(\omega)})^2}
 \eeqn
is a function of $x=\rho/\rho_0$, with the vector density
$\rho=\sqrt{j_\mu j^\mu}$,
$j_\mu=\sum_B\bar{\psi}_B\gamma_\mu\psi_B$ and the saturation
density of nuclear matter $\rho_0$. The eight real parameters in
Eq. (\ref{fx}) are not independent. The five constraints
$f_{\sigma(\omega)}(1)=1$, $f''_{\sigma(\omega)}(0)=0$ and
$f''_\sigma(1)=f''_\omega(1)$ reduce the number of independent
parameters to three. The density-dependent $\rho$-meson coupling
constant $g_{\rho B}$ is introduced as,
 \beqn
 \label{grho}
  g_{\rho B}(\rho)=g_{\rho B}(\rho_0)\exp[-a_\rho(x-1)],
 \eeqn
with two parameters $a_\rho$ and $g_{\rho B}(\rho_0)$.

The equations of motion for baryons and mesons can be derived from
the Lagrangian density in Eq. (\ref{lb}). In the following, we
present only the case for the density-dependent couplings. The
equations for nonlinear couplings can be obtained easily by adding
the nonlinear self-couplings of the mesons and neglecting the
density dependencies of the coupling constants.

The equations of motion for baryons are:
 \beqn
  \left[\gamma^\mu(i\partial_\mu-g_{\omega B}\omega_\mu-g_{\rho B}\btau_B\cdot\brho_\mu
    -e\frac{1-\tau_{3B}}{2}A_\mu-\Sigma_{\mu B}^R)-m_B^*\right]\psi_B=0,
 \eeqn
where the effective mass $m_B^*=m_B+g_{\sigma B}\sigma$ and
$\Sigma_{\mu B}^R$ is a ``rearrangement" term due to the density
dependencies of the couplings:
 \beqn
  \Sigma_{\mu B}^R=\frac{j_\mu}{\rho}(\frac{\partial
                g_{\omega B}}{\partial\rho}\bar{\psi}_B\gamma^\nu\psi_B\omega_\nu
              +\frac{\partial
                g_{\rho B}}{\partial\rho}\bar{\psi}_B\gamma^\nu\btau_B\psi_B\brho_\nu
              +\frac{\partial
              g_{\sigma B}}{\partial\rho}\bar{\psi}_B\psi_B\sigma).
 \eeqn
The field equations for mesons and photons are respectively:
 \beqn
 \label{sigma0}
  (\partial^\mu\partial_\mu+m_\sigma^2)\sigma&=&-\sum_Bg_{\sigma B}\bar{\psi}_B\psi_B,\\
 \label{omega0}
   \partial_\mu\omega^{\mu\nu}+m_\omega^2\omega^\nu&=&\sum_Bg_{\omega B}\bar{\psi}_B\gamma^\nu\psi_B,\\
 \label{rho0}
   \partial_\mu\brho^{\mu\nu}+m_\rho^2\brho^\nu&=&\sum_Bg_{\rho B}
        \bar{\psi}_B\gamma^\nu\btau_B\psi_B+g_{\rho B}\brho_\mu\times \brho^{\mu\nu},\\
   \partial_\mu
   A^{\mu\nu}&=&e\bar{\psi}\gamma^\nu\frac{1-\tau_{3B}}{2}\psi.
 \eeqn
\subsection{Nuclear Matter}
For infinite nuclear matter, introducing the mean-field
approximation, i.e., the meson fields are replaced by their mean
values, and neglecting the coulomb field, the baryon wave function
is the eigenstate of momentum $k$, and the source currents
$\bar{\psi}_B\psi_B$ and $\bar{\psi}_B\gamma^\nu\psi_B$ in Eqs.
(\ref{sigma0}-\ref{rho0}) are independent of the spatial
coordinate x. Thus the equations of motion can be simplified as:

 \beqn
  \label{baryon0}
    &&\left[\gamma^\mu(k_\mu-g_{\omega B}\omega_\mu-g_{\rho B}{\btau_B}\cdot\brho_\mu
       -\Sigma_{\mu B}^R)-m_B^*\right]\psi_B(k)=0,\\
  \label{sigma}
    &&m_\sigma^2\sigma=-g_{\sigma B}\rho_{s},\\
  \label{omega}
    &&m_\omega^2\omega_0=g_{\omega B}\rho,\\
  \label{rho}
    &&m_\rho^2\rho_{0,3}=\sum_B g_{\rho B}\tau_{3B}\rho_B.
 \eeqn
For the nonlinear self-coupling effective interactions, the
corresponding terms $-U'(\sigma)$ and $-U'(\omega_0)$ should be
taken into account in the Eqs. (\ref{sigma}) and (\ref{omega})
respectively.

The eigenvalues of the Dirac equation for baryons in Eq.
(\ref{baryon0}) are obtained as:
 \beqn
 \label{values}
  e_B(k)=g_{\omega B}\omega_0+g_{\rho B}\cdot\tau_{3B}\cdot\rho_{0,3}+\Sigma_{0B}^R+\sqrt{k^2+m_B^{*2}},
 \eeqn
where $\Sigma_{0B}^R$ is the time component of the rearrangement
term. The baryon vector density $\rho$ and scalar density $\rho_s$
are respectively:
 \beqn
    \rho&=&\sum_B<\bar{\psi}_B\gamma^0\psi_B>=\sum_B\rho_B=\sum_B\frac{k_B^3}{3\pi^2},\\
  \label{rhos}
    \rho_s&=&\sum_B<\bar{\psi}_B\psi_B>=\sum_B\rho_{sB}
      =\frac{1}{\pi^2}\sum_B\int_0^{k_B}k^2dk\frac{m_B^*}{\sqrt{k^2+m_B^{*2}}},
 \eeqn
where $k_B$ denotes the Fermi momentum of baryon $B$, and the
no-sea approximation has been used.

The energy density and pressure of nuclear matter are
respectively,
 \beqn
   \label{energy}
    \varepsilon&=&\frac{1}{2}m_\sigma^2\sigma^2+\frac{1}{2}m_\omega^2\omega_0^2
       +\frac{1}{2}m_\rho^2\rho_{0,3}^2
       +\frac{1}{\pi^2}\sum_B\int_0^{k_B}k^2dk\sqrt{k^2+m_B^{*2}},\\
   \label{pressure}
     P&=&-\frac{1}{2}m_\sigma^2\sigma^2+\frac{1}{2}m_\omega^2\omega_0^2
         +\frac{1}{2}m_\rho^2\rho_{0,3}^2+\sum_B\rho_B\cdot\Sigma_{0B}^R
         +\frac{1}{3\pi^2}\sum_B\int_0^{k_B}\frac{k^4}{\sqrt{k^2+m_B^{*2}}}dk.
 \eeqn


\subsection{Neutron Stars}
The charge-neutral neutron stars includes not only neutrons and
protons, but also leptons $\lambda$ (mainly $e^-$ and $\mu^-$ ) in
equal number to protons and also hyperons at high densities. The
Lagrangian density for neutron stars is similar to Eq. (\ref{lb}),
except an additional term for leptons:
 \beqn
   \label{lbs}
   {\cal L}_\lambda=\sum_{\lambda=e^-,\mu^-}\bar{\psi}_\lambda
       (i\gamma^\mu\partial_\mu-m_\lambda)\psi_\lambda,
 \eeqn

Introducing the mean-field and no-sea approximation, the equations
of motion for baryons and mesons can be derived, and the
corresponding energy eigenvalues, baryon density and scalar
density can be obtained for neutron stars, similar to those for
the nuclear matter. The equations of motion for electron and
$\mu^-$ are free Dirac equations and their densities can be
expressed in terms of their corresponding Fermi momenta as
$\rho_\lambda=k_\lambda^3/(3\pi^2)$.

The chemical potentials $\mu_B$ for the baryons $B$ are the energy
eigenvalues of the Dirac equation: $\mu_B=\varepsilon_B(k)$. The
chemical potentials $\mu_\lambda$ for the leptons are the
solutions of their equations of motion:
$\mu_\lambda=\sqrt{k_\lambda^2+m_\lambda^2}$. In neutron stars,
the chemical equilibrium conditions are:
 \beqn
 \label{chemical}
  \mu_B=b_B\mu_n-q_B\mu_e,~~~\mu_\mu=\mu_e,
 \eeqn
where $b_B$ and $q_B$ denote baryon charge and electronic charge
of baryon B, $\mu_n$, $\mu_e$ and $\mu_\mu$ are the chemical
potentials for neutron, electron and $\mu^-$ respectively. The
baryon number conservation and charge-neutral conditions are given
by:
 \beqn
  \label{baryon}
   \rho&=&\sum_B\rho_B=\sum_B\frac{b_Bk_B^3}{3\pi^2},\\
  \label{charge} Q&=&\sum_BQ_B+\sum_\lambda Q_\lambda=\sum_B
   \frac{q_Bk_B^3}{3\pi^2}-\sum_\lambda\frac{k_\lambda^3}{3\pi^2}=0.
 \eeqn

The energy density and pressure for neutron stars are
respectively:
 \beqn
  \label{energy1}
  \varepsilon&=&\frac{1}{2}m_\sigma^2\sigma^2+\frac{1}{2}m_\omega^2\omega_0^2
             +\frac{1}{2}m_\rho^2\rho_{0,3}^2\nonumber\\
         &&+\frac{1}{\pi^2}\sum_B\int_0^{k_B}k^2dk\sqrt{k^2+m_B^{*2}}
           +\frac{1}{\pi^2}\sum_{\lambda=e^-,\mu^-}\int_0^{k_\lambda}k^2dk{\sqrt{k^2+m_\lambda^2}}~~,\\
  \label{pressure1}
  P&=&-\frac{1}{2}m_\sigma^2\sigma^2+\frac{1}{2}m_\omega^2\omega_0^2
         +\frac{1}{2}m_\rho^2\rho_{0,3}^2+\sum_B\rho_B\cdot\Sigma_{0B}^R\nonumber\\
    &&+\frac{1}{3\pi^2}\sum_B\int_0^{k_B}\frac{k^4}{\sqrt{k^2+m_B^{*2}}}dk
      +\frac{1}{3\pi^2}\sum_{\lambda=e^-,\mu^-}\int_0^{k_\lambda}dk\frac{k^4}{\sqrt{k^2+m_\lambda^2}}~.
 \eeqn


\section{Results and Discussions}

For nuclear matter, Eqs. (\ref{sigma})-(\ref{rhos}) provide a set
of coupled transcendental relations defining the meson fields and
energy eigenvalues. The list of unknowns is
 \beqn
 \rho,~~~k_B,~~~\rho_s,~~~g_{\omega B},~~~ g_{\sigma B},~~~ g_{\rho B},~~~
 \sigma,~~~\omega_0,~~~\rho_{0,3},~~~\varepsilon_B(k).\nonumber
 \eeqn
For a given baryon density $\rho$ and the asymmetry of the nuclear
matter $t=(\rho_n-\rho_p)/\rho$, we can get the coupling constants
$g_{\sigma B}$, $g_{\omega B}$ and $g_{\rho B}$ from Eqs.
(\ref{grho0})-(\ref{grho}), as well as the neutron density
$\rho_n$ and proton density $\rho_p$, and of course their
corresponding Fermi momenta $k_n$ and $k_p$. The $\omega$ and
$\rho$ fields can be obtained by Eqs. (\ref{omega}) and
(\ref{rho}) and the $\sigma$ field can be solved from Eqs.
(\ref{sigma}) and (\ref{rhos}) by iteration.

The properties of neutron stars can be obtained by solving the
Eqs. (\ref{sigma})-(\ref{rhos}) and
(\ref{chemical})-(\ref{charge}) by the following procedures: (1)
For a given baryon density $\rho$, taking initial values of the
meson fields ($\sigma$, $\omega_0$, $\rho_{0,3}$) as well as the
neutron and electron chemical potentials ($\mu_n,~~\mu_e$), the
particle densities and Fermi momenta for electrons
($k_e,~~\rho_e$), protons ($k_p,~~\rho_p$) and hyperons
($k_h,~~\rho_h$) can be obtained via the eigenvalues in Eq.
(\ref{values}), chemical equilibriums and charge-neutral
conditions in Eqs. (\ref{chemical}) and (\ref{charge}), which in
turn fix the neutron density $\rho_n$ from baryon number
conservations Eq. (\ref{baryon}); (2) With the particle densities
($\rho_B$, $\rho_e$, and $\rho_\mu$) and Fermi momenta ($k_B$,
$k_e$, and $k_\mu$), the meson fields ($\sigma$, $\omega_0$,
$\rho_{0,3}$) can be obtained from Eqs. (\ref{sigma})-(\ref{rho});
(3) These two steps should be repeated by iteration until the
self-consistence is achieved.


\subsection{Effective interaction strengths in nuclear matter and neutron stars}

Using the nonlinear RMF interactions  NL1, NL2 \cite{joon86}, NL3
\cite{lala97}, NLSH \cite{shar93}, TM1, TM2 \cite{suga94}, and
GL-97 \cite{glen97} and the density-dependent interactions TW-99
\cite{type99}, and DD-ME1 \cite{niks02}, the density dependencies
of various effective interaction strengths in RMF theory are
studied and carefully compared in nuclear matter and neutron
stars.

\begin{figure}[ht]
 \centering
 \includegraphics[scale=0.5]{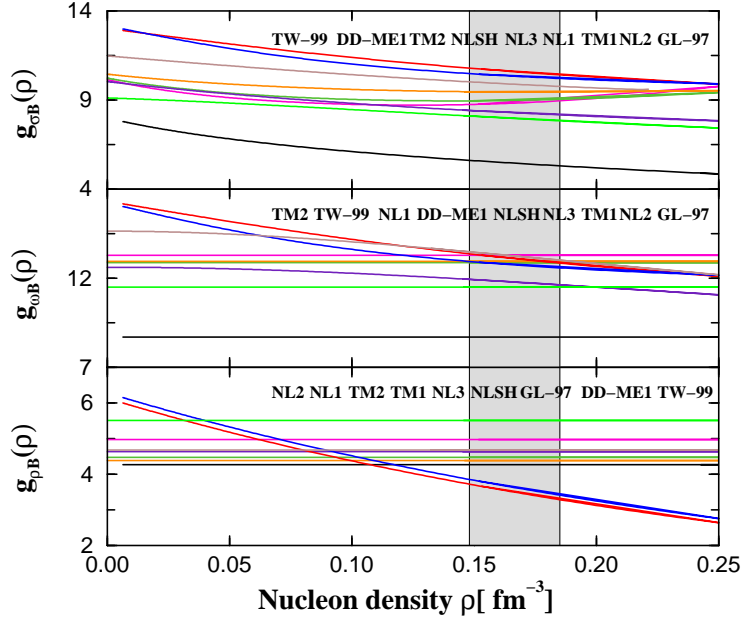}
 \caption{(Color online) The effective interaction strengths for
 $\sigma$ (top), $\omega$ (middle) and $\rho$ (bottom) in symmetric
 nuclear matter as functions of the
nucleon density. The shadowed area corresponds to the empirical
value of the saturation density in nuclear matter ( Fermi momentum
$k_F=1.35\pm0.05$ fm$^{-1}$ or density $\rho=0.166\pm0.018$
fm$^{-3}$). The curves are labelled from the top to the bottom at
$\rho$=0.15 fm$^{-3}$ orderly from left to right.}
 \label{fig:fig1}
\end{figure}

In Fig. \ref{fig:fig1}, the density dependencies of the effective
interaction strengths for $\sigma$ (top), $\omega$ (middle) and
$\rho$ (bottom) mesons in symmetric nuclear matter as functions of
the nucleon density are shown. The shadowed area corresponds to
the empirical value of the saturation density in nuclear matter (
Fermi momentum $k_F=1.35\pm0.05$ fm$^{-1}$ or density
$\rho=0.166\pm0.018$ fm$^{-3}$). These curves are labelled from
the top to the bottom at $\rho$=0.15 fm$^{-3}$ orderly from left
to right. For the nonlinear effective interaction, the
``equivalent" density dependencies of the effective interaction
strengths for $\sigma$, $\omega$ and $\rho$ are extracted from the
meson field equations Eqs. (\ref{sigma}) - (\ref{rho}) according
to:
 \beqn
  \label{sigma1}
  g_{\sigma B}(\rho)&=&g_{\sigma B}+U'(\sigma)/\rho_s=g_{\sigma B}+(g_2\sigma^2+g_3\sigma^3)/\rho_s,\\
  g_{\omega B}(\rho)&=&g_{\omega B}-U'(\omega_0)/\rho=g_{\omega B}-(c_3\omega_0^3)/\rho,\\
  g_{\rho B}(\rho)&=&g_{\rho B}.
\eeqn

The density dependencies of the interaction strengths for TW-99
and DD-ME1 are very similar for symmetric nuclear matter in Fig.
\ref{fig:fig1}, as noted in Ref. \cite{niks02}. Here the
comparison between the nonlinear and the density-dependent
interaction will be emphasized.

For the $\sigma$-meson, the interaction strengths given by TW-99
and DD-ME1 are quite different from the others in either
magnitudes or slopes. In particular, strengths of TW-99 and DD-ME1
for the density interval in Fig. \ref{fig:fig1} are almost twice
as large as that of GL-97. Differences for nonlinear and
density-dependent interactions can also be seen in the region of
the empirical nuclear matter densities. For the $\omega$-meson,
except TW-99, DD-ME1, TM1 and TM2, all the other strengths are
density-independent. All the strengths are similar to each other
in the region of the empirical saturation densities compared with
those of the $\sigma$-meson, although large differences can also
be seen at low densities. For the $\rho$-meson which describes the
isospin asymmetry, the strengths for TW-99 and DD-ME1 show strong
density dependencies in contrast with the constants in the other
interactions. They cross the nonlinear interactions at a density
much lower than the empirical saturation density.

\begin{figure}[ht]
 \centering
 \includegraphics[scale=0.5]{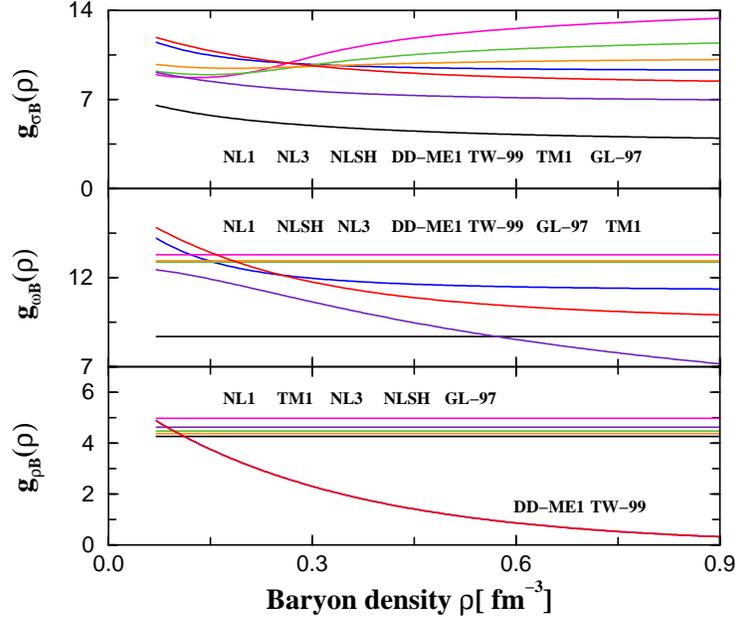}
 \caption{(Color online) Similar as Fig. \ref{fig:fig1}, but for
neutron stars matter. The curves are labelled from the top to the
bottom at $\rho$=0.9 fm$^{-3}$ orderly from left to right.}
 \label{fig:fig2}
\end{figure}

The interaction strengths as functions of baryon density for
neutron stars matter are given in Fig. \ref{fig:fig2}. As the
effective interactions NL2 and TM2 are mainly used in light
nuclei, we don't discuss them here. At densities $\rho<0.2$
fm$^{-3}$, Fig. \ref{fig:fig2} is similar to Fig. \ref{fig:fig1}.
For the scalar $\sigma$-meson, the interaction strengths of TW-99,
DD-ME1, TM1 and GL-97 decrease with the baryon density in similar
slopes, while those of NL1, NL3 and NLSH decrease with baryon
density for the densities $\rho<0.2$ fm$^{-3}$, then increase
afterwards. This is due to the positive $g_3\sigma^3$ in Eq.
(\ref{sigma1}) for NL1, NL3 and NLSH, in contrast with the
negative ones for TM1 and GL-97. For the vector $\omega$-meson,
the interaction strengths of TW-99, DD-ME1 and TM1 decrease with
the baryon density. At densities $0.2 <\rho<0.55$ fm$^{-3}$, the
strengths of DD-ME1, TW-99 and TM1 are between those of NL3 and
GL-97. The curve given by TM1 crosses with the line of GL-97 at
density $\rho\approx0.57$ fm$^{-3}$, and then gives the weakest
interaction strength. For isospin-vector $\rho$ meson, the
interaction strengths of TW-99 and DD-ME1 decrease with baryon
density and trend to vanish at high densities, while the others
are constants.

Although the aspects of the interaction strengths are quite
different from each other, as it will be shown in the following
sections, all of them give fair descriptions for the properties of
nuclear matter.


\subsection{Potentials for nuclear matter and neutron stars}

\begin{figure}[ht]
 \centering
 \includegraphics[scale=0.5]{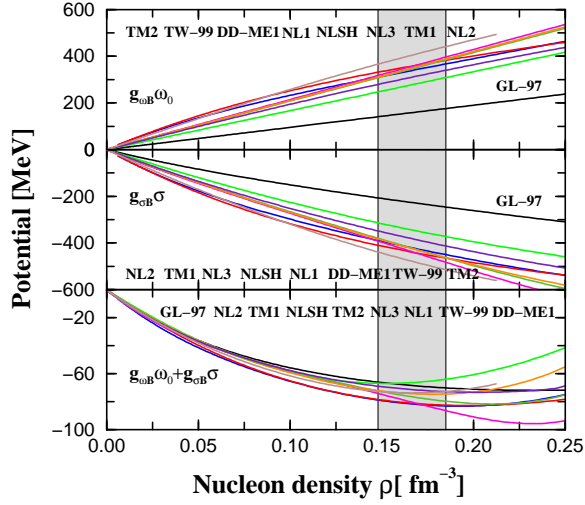}
 \caption{(Color online) Vector potentials $g_{\omega B}\omega_0$
(top), scalar potentials $g_{\sigma B}\sigma$  (middle) and the
sum of both (bottom) in symmetric nuclear matter for different
effective interactions (as marked in the figure) as functions of
the nucleon density $\rho$. The shadowed area corresponds to the
empirical value of the saturation density in nuclear matter. The
curves are labelled from the top to the bottom at $\rho$=0.15
fm$^{-3}$ orderly from left to right.}
 \label{fig:fig3}
\end{figure}

\begin{figure}[ht]
 \centering
 \includegraphics[scale=0.5]{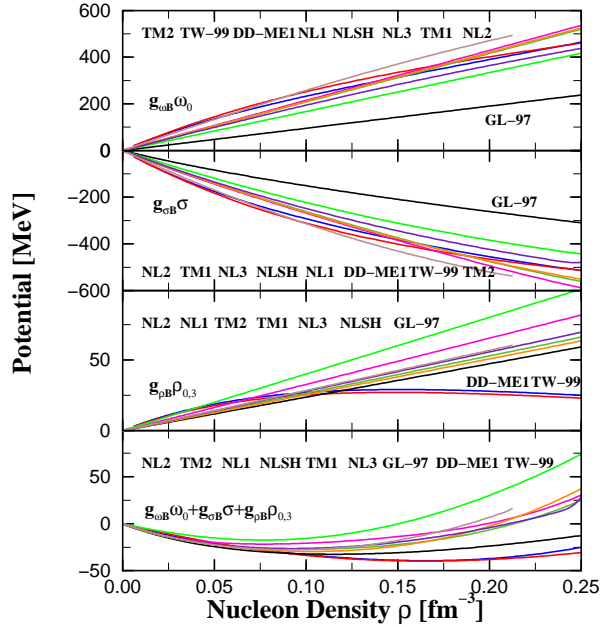}
 \caption{(Color online) Vector potentials $g_{\omega B}\omega_0$,
scalar potentials $g_{\sigma B}\sigma$, isospin vector potentials
$g_{\rho B}\rho_{0,3}$ and the sum of them in pure neutron matter
for different effective interactions as functions of the nucleon
density $\rho$. The curves are labelled from the top to the bottom
at $\rho$=0.15 fm$^{-3}$ orderly from left to right.}
 \label{fig:fig4}
\end{figure}

\begin{figure}[ht]
 \centering
 \includegraphics[scale=0.5]{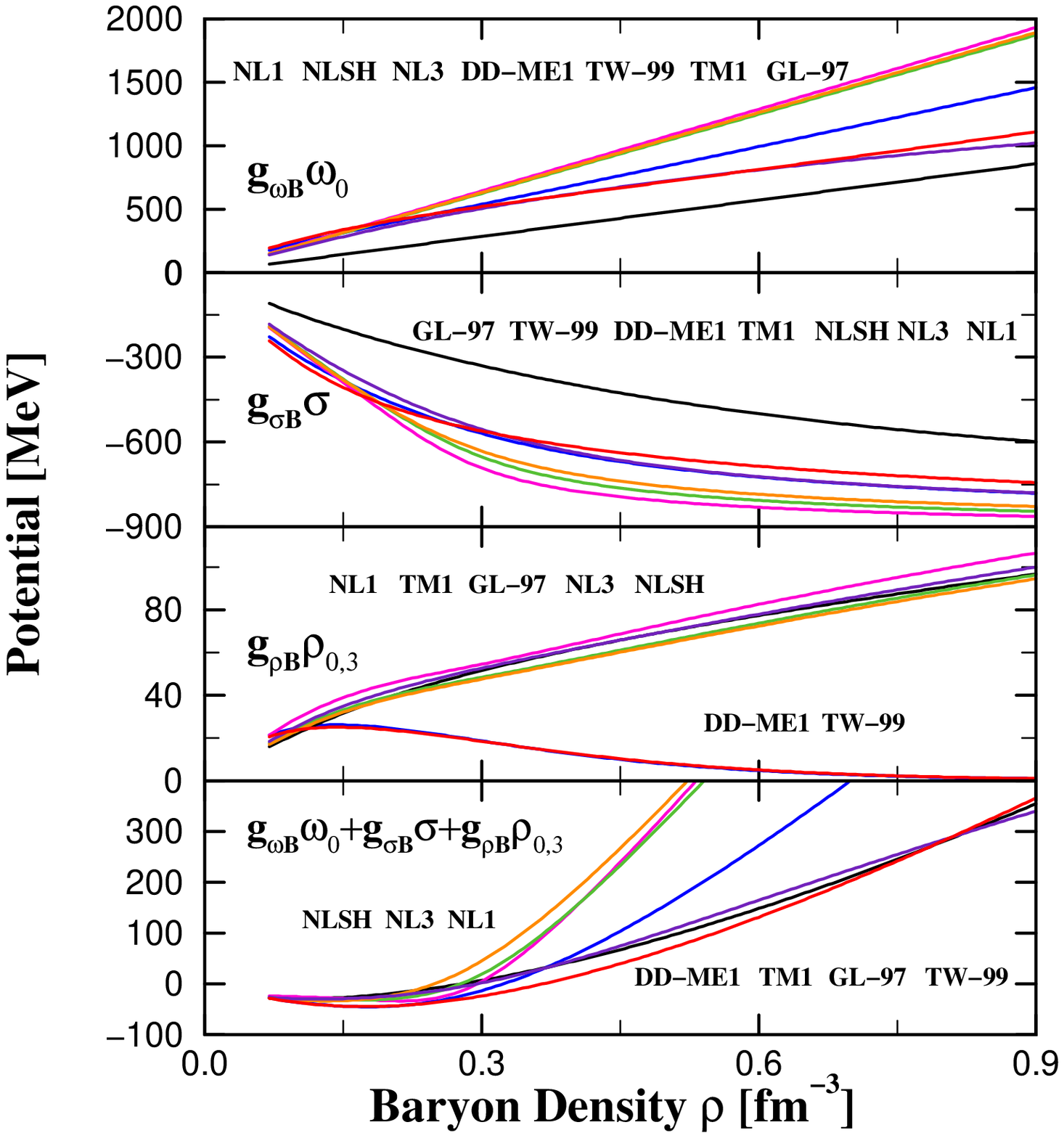}
 \caption{(Color online) Vector potentials $g_{\omega B}\omega_0$,
scalar potentials $g_{\sigma B}\sigma$, isospin vector potentials
$g_{\rho B}\rho_{0,3}$ and the sum of them in neutron stars with
neutrons and protons only, for different effective interactions as
functions of the baryon density $\rho$. The curves are labelled
from the top to the bottom at $\rho$=0.45 fm$^{-3}$ orderly from
left to right.}
 \label{fig:fig5}
\end{figure}

Potentials for symmetric nuclear matter, pure neutron matter and
neutron stars calculated with density-dependent interactions TW-99
and DD-ME1 are illustrated in Figs. \ref{fig:fig3},
\ref{fig:fig4}, and \ref{fig:fig5}. The results are shown in
comparison with those obtained with the nonlinear interactions,
such as NL1, NL2, NL3, NLSH, TM1, TM2 and GL-97.

As the contribution of $\rho$-meson potential for symmetric
nuclear matter vanishes, we show the vector potentials $g_{\omega
B}\omega_0$, scalar potentials $g_{\sigma B}\sigma$ and their sum
as functions of nucleon density $\rho$ for different effective
interactions which are marked from top to bottom at density
$\rho$=0.15 fm$^{-3}$ orderly from left to right in Fig.
\ref{fig:fig3}. For the vector and scalar potentials, GL-97 gives
the weakest results due to its weakest interaction strength in
Fig. \ref{fig:fig1}. The curves associated with the other
interactions lie between those by TM2 and NL2. Their difference
increases with the density, the difference of scalar potentials at
$\rho$=0.15 fm$^{-3}$ between NL2 and TM2 is around 120 MeV, and
that of vector potentials is around 130 MeV. At saturation
density, the scalar and vector potentials given by the
density-dependent interactions TW-99 and DD-ME1 are similar to the
nonlinear interactions except for GL-97, TM2 and NL2. For the
total potentials for symmetric nuclear matter $g_{\omega
B}\omega_0$+$g_{\sigma B}\sigma$, GL-97 also gives the weakest
result and the difference for different interactions is about
20$\sim$30 MeV in the range of saturated densities.

The vector potentials $g_{\omega B}\omega_0$, scalar potentials
$g_{\sigma B}\sigma$, isospin potentials $g_{\rho B}\rho_{0,3}$
and their sum as functions of the nucleon density for different
effective interactions are shown in Fig. \ref{fig:fig4} for pure
neutron matter. The variations of the vector potentials with
density $\rho$ are same as those for symmetric nuclear matter
because they are related to the whole nucleon density $\rho$ only.
The variations of the scalar potentials with density $\rho$ are
slightly different from those for symmetric nuclear matter, as the
same density for symmetric nuclear matter and pure neutron matter
implicates different Fermi momentum, $k_B$, as shown in Eq.
(\ref{rhos}). The $\rho$-meson provides the necessary isospin
asymmetry. In pure neutron matter, the isospin potentials $g_{\rho
B}\rho_{0,3}$ given by nonlinear interactions increase linearly
with the density due to their constant interaction strengths in
Fig. \ref{fig:fig1}. While for the same reasons, we can easily
understand the isospin potentials for the density-dependent
interactions. The compensation between the density dependencies of
$g_{\rho B}$ and the increase of the density makes the isospin
potentials increase at first and decrease after $\rho=0.12$
fm$^{-3}$. The sum of the vector potentials $g_{\omega
B}\omega_0$, scalar potentials $g_{\sigma B}\sigma$, and isospin
potentials $g_{\rho B}\rho_{0,3}$ gives the total potentials for
baryon in pure neutron matter. The total potentials are attractive
at low densities and become repulsive at high densities. Different
from those of symmetric nuclear matter, due to the contribution of
the isospin potentials for pure neutron matter, the difference
between the total potentials is quite large, e.g., the largest
difference 40 MeV between NL2 and TW-99 occurs at $\rho=$ 0.15
fm$^{-3}$.

For neutron stars, Fig. \ref{fig:fig5} shows the vector potentials
$g_{\omega B}\omega_0$, scalar potentials $g_{\sigma B}\sigma$,
iso-vector potentials $g_{\rho B}\rho_{0,3}$ and the sum of them
for different effective interactions as functions of the baryon
density $\rho$. For scalar and vector potentials at densities
$\rho<0.2$ fm$^{-3}$, the properties are similar to those in Figs.
\ref{fig:fig3} and \ref{fig:fig4}. The vector potentials
$g_{\omega B}\omega_0$, which offer short-range repulsive
interactions, increase with the baryon density, while the scalar
potentials $g_{\sigma B}\sigma$, which offer medium-range
attractive interactions, change evidently at densities $\rho<0.3$
fm$^{-3}$, and reach saturation at density $\rho\approx0.4$
fm$^{-3}$. Just as for nuclear matter, GL-97 always gives the
weakest scalar and vector potentials. Due to the strong
interaction strengths of NL1, NL3 and NLSH at high densities as
shown in Fig. \ref{fig:fig2}, they give large scalar and vector
potentials at densities $\rho>0.3$ fm$^{-3}$. The results
calculated with DD-ME1, TW-99 and TM1 lie in the middle. The
difference in potentials between the two density-dependent
effective interactions DD-ME1 and TW-99 shows up clearly beyond
the density $\rho=$0.4 fm$^{-3}$, in contrast with that for
nuclear matter.

As shown in Fig. \ref{fig:fig5}, the contribution of the
$\rho$-meson provides the necessary isospin asymmetry for neutron
stars. For densities $0.065<\rho<$0.1 fm$^{-3}$, the potential
$g_{\rho B}\rho_{0,3}$ calculated with various interactions range
from 20 MeV to 30 MeV. For nonlinear interactions, the potentials
$g_{\rho B}\rho_{0,3}$ increase with the baryon density and are
about 100 MeV at density $\rho$=0.9 fm$^{-3}$, which are about
10\% of the vector potentials $g_{\omega B}\omega_0$ for GL-97 and
TM1, and 5\% of those for NL1, NL3 and NLSH. For density-dependent
interactions, the isospin potentials trend to vanish after
$\rho$=0.75 fm$^{-3}$, due to the density dependencies of the
$\rho$-meson interaction strengths in Fig. \ref{fig:fig2}.

The total potentials $g_{\omega B}\omega_0$+$g_{\sigma
B}\sigma+g_{\rho B}\rho_{0,3}$ are attractive at low densities. At
high densities, they become repulsive and the neutron stars are
bound by the gravity. The results given by various interactions
are similar at densities $\rho<$0.2 fm$^{-3}$ and become quite
different at $\rho>$0.3 fm$^{-3}$. The total potentials for NL1,
NL3 and NLSH are close to each other and increase rapidly. Those
for TW-99, GL-97 and TM1 are close to each other after
$\rho\simeq$ 0.16 fm$^{-3}$, while that of DD-ME1 lies in-between.


\subsection{Properties of nuclear matter}

\begin{figure}
 \centering
 \includegraphics[scale=0.8]{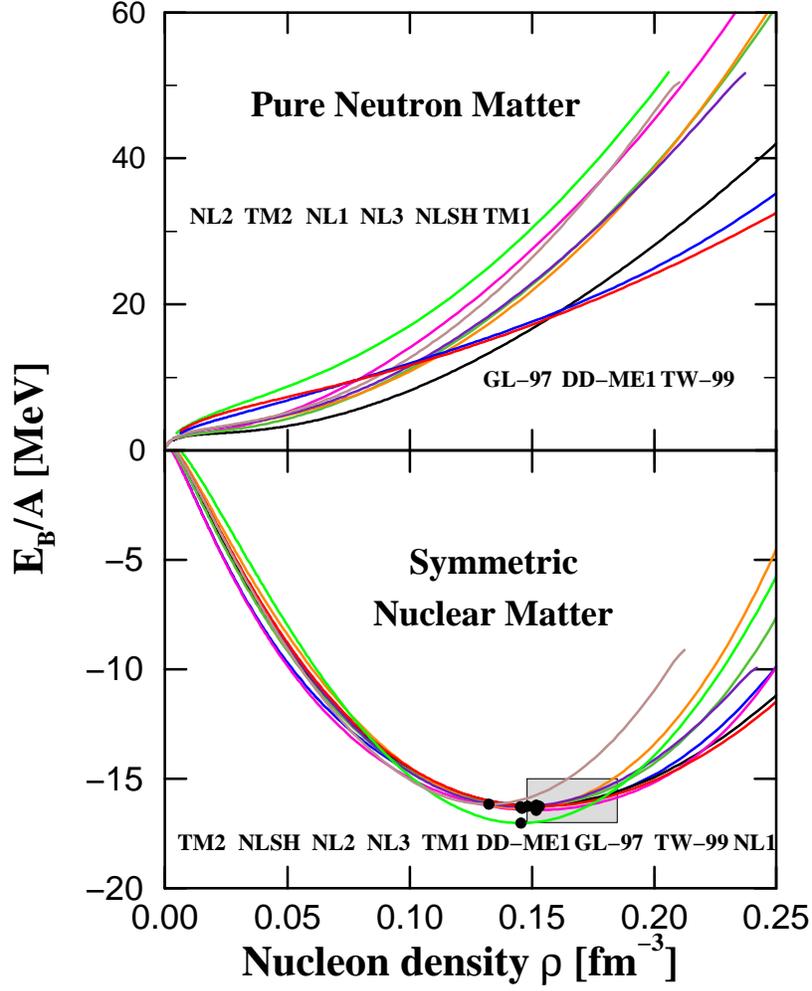}
 \caption{(Color online) Energies per nucleon,
$E_B/A=\varepsilon$/$\rho-m$, in pure neutron matter (upper panel)
and symmetric nuclear matter (lower panel) for difference
effective interactions as functions of nucleon density $\rho$. The
dots in the lower panel correspond to the saturation densities in
symmetric nuclear matter and the shadowed area corresponds to the
empirical value of saturation densities in symmetric nuclear
matter (density $\rho=0.166\pm0.018$ fm$^{-3}$ and energy per
particle $\varepsilon/\rho=-16.0\pm1.0$MeV). The curves are
labelled from the top to the bottom at $\rho$=0.2 fm$^{-3}$
orderly from left to right.}
 \label{fig:fig6}
\end{figure}

In Fig. \ref{fig:fig6} we display the energies per nucleon,
$E_B/A=\varepsilon$/$\rho-m$, for pure neutron matter (upper
panel) and symmetric nuclear matter (lower panel) for
density-dependent interactions TW-99 and DD-ME1 as functions of
nucleon density $\rho$. For comparison, the results for various
nonlinear interactions are also shown. We label the curves from
top to bottom at density $\rho$=0.20 fm$^{-3}$ orderly from left
to right.

In the upper panel of Fig. \ref{fig:fig6}, the energies per
nucleon for all interactions are always positive and increase with
the nucleon density for pure neutron matter. The results for
density-dependent interactions DD-ME1 and TW-99 are very similar
to each other. At densities $\rho<$0.075 fm$^{-3}$,  they are
larger than the other interactions except for NL2, thereafter they
cross the curves for the other interactions and give the smallest
energy per nucleon after densities $\rho>$0.17 fm$^{-3}$.

For symmetric nuclear matter, the curves for energies per nucleon
for various interactions display similar dependencies on densities
below the saturation density and pronounced differences at higher
densities. The saturation densities for various interactions are
similar ($\rho=0.166\pm0.018$ fm$^{-1}$, $E_B/A=-16.33\pm0.10$
MeV) and are located in the empirical range of saturation density
except those of TM2 and NL2, which are mainly used for light
nuclei.

From the upper and lower panels of Fig. \ref{fig:fig6}, it can be
seen that the equations of state at low densities (below
saturation density) for pure neutron matter and symmetric nuclear
matter are different. The deviations of energy per nucleon $E_B/A$
for symmetric nuclear matter below saturation density are
negligible compared with those for pure neutron matter (the
deviation is about 15 MeV at density $\rho=0.15$ fm$^{-3}$). The
different results for symmetric nuclear matter and pure neutron
matter mainly come from the potentials as shown in Figs.
\ref{fig:fig3} and \ref{fig:fig4}, especially from the isospin
potentials $g_{\rho B}\rho_{0,3}$, which exhibits large deviation.

From Fig. \ref{fig:fig3}, we have seen that the difference for the
total potentials at saturated density in symmetric nuclear matter
is about 20$\sim$30 MeV, while the energies per nucleon at
saturation density are close to each other in Fig. \ref{fig:fig6}.
To understand these, we will discuss energy per nucleon at the
saturation density for symmetric nuclear matter in detail.

From Eq. (\ref{energy}) and the meson field equations Eqs.
(\ref{sigma}) - (\ref{rho}), the energy density for symmetric
nuclear matter can be expressed as:
 \beqn
 \label{enegy2}
 \varepsilon&=&\frac{1}{2}m_\sigma^2\sigma^2+U(\sigma)
         -\frac{1}{2}m_\omega^2\omega_0^2-U(\omega_0)- \sum_B\rho_B\cdot\Sigma_{0B}^R\nonumber\\
       &&+[m_\omega^2\omega_0^2+4\cdot U(\omega_0)+\sum_B\rho_B\cdot\Sigma_{0B}^R
         +\frac{1}{\pi^2}\sum_B\int_0^{k_B}k^2dk\sqrt{k^2+(m_B+g_{\sigma B}\sigma)^2}~]\nonumber\\
       &=&-\frac{1}{2}g_{\sigma B}\sigma\rho_s-\frac{1}{6}g_2\sigma^3-\frac{1}{4}g_3\sigma^4
          -\frac{1}{2}g_{\omega B}\omega_0\rho+\frac{1}{4}c_3\omega_0^4-\sum_B\rho_B\cdot\Sigma_{0B}^R\nonumber\\
        &&+[g_{\omega B}\omega_0\rho+\sum_B\rho_B\cdot\Sigma_{0B}^R
          +\frac{1}{\pi^2}\sum_B\int_0^{k_B}k^2dk\sqrt{k^2+(m_B+g_{\sigma B}\sigma)^2}~]\nonumber\\
       &=&\varepsilon_\sigma+\varepsilon_\omega+\varepsilon_{re}+\varepsilon_N,
 \eeqn
 where,
 \beqn
  \varepsilon_\sigma&=&-\frac{1}{2}g_{\sigma B}\sigma\rho_s-\frac{1}{6}g_2\sigma^3-\frac{1}{4}g_3\sigma^4,\\
  \varepsilon_\omega&=&-\frac{1}{2}g_{\omega B}\omega_0\rho+\frac{1}{4}c_3\omega_0^4,\\
  \varepsilon_{re}&=&-\sum_B\rho_B\cdot\Sigma_{0B}^R
  ,\\
  \varepsilon_N^p&=&g_{\omega B}\omega_0\rho+\sum_B\rho_B\cdot\Sigma_{0B}^R,\\
  \varepsilon_N^k&=&\frac{1}{\pi^2}\sum_B\int_0^{k_B}k^2dk\sqrt{k^2+(m_B+g_{\sigma B}\sigma)^2},\\
  \label{enegyn}
  \varepsilon_N&=&\varepsilon_N^p+\varepsilon_N^k,
 \eeqn
in which, the contributions from $\sigma$ and $\omega$ fields are
respectively $\varepsilon_\sigma$ and $\varepsilon_\omega$, the
rearrangement term is $\varepsilon_{re}$, and the contribution
from nucleons is $\varepsilon_N$ with its potential energy part
$\varepsilon_N^p$ and kinetic energy part $\varepsilon_N^k$. For
nonlinear interactions, $\varepsilon_{re}$=0, while for the
density-dependent interactions, $g_2=g_3=c_3=0$.

In Table \ref{tab:table1}, the Fermi momenta $k_F$
($\rho_0={k_F^3}/{3\pi^2}$),  vector densities $\rho_0$, scalar
densities $\rho_s$, scalar potentials $g_{\sigma B}\sigma$, vector
potentials $g_{\omega B}\omega_0$, the various contributions
$\varepsilon_\sigma$, $\varepsilon_\omega$, $\varepsilon_{re}$,
$\varepsilon_N^p$, $\varepsilon_N^k$, and $\varepsilon_N$, the
system energy densities $\varepsilon$, nucleon mass $m$ and
energies per nucleon $E_B/A=\varepsilon/\rho-m$ at saturation
density for different interactions are shown. The saturation
densities given by various interactions are similar
($\rho_0\approx$0.150 fm$^{-3}$), except for TM2 ($\rho_0=$0.132
fm$^{-3}$). From Eqs. (\ref{enegy2}) - (\ref{enegyn}), we know
that if the scalar densities $\rho_s$ and vector densities
$\rho_0$ (or Fermi momenta $k_F$) at saturation densities are
similar, larger scalar potentials (negative) will give larger
$\varepsilon_\sigma$ (positive) and smaller $\varepsilon_N^k$
(positive), and larger vector potentials will give larger
$\varepsilon_\omega$ (negative) and $\varepsilon_N^p$ (positive).
Taking into account the contribution of the rearrangement term
$\varepsilon_{re}$, the energies per nucleon
$E_B/A=\varepsilon/\rho-m$ for various interactions become closer
to each other. For example, for TW-99 and GL-97, which give the
largest and smallest potentials respectively, the difference of
$g_{\sigma B}\sigma\sim$ -210.5 MeV leads to $\Delta
\varepsilon_\sigma=$16.04 MeV fm$^{-3}$ and $\Delta
\varepsilon_N^k=$-30.80 MeV fm$^{-3}$, and the difference of
$g_{\omega B}\omega\sim$ 194.1 MeV leads to $\Delta
\varepsilon_\omega=$-14.77 MeV fm$^{-3}$ and $\Delta
\varepsilon_N^p=$29.70 MeV fm$^{-3}$. Taking into account the
contribution of the rearrangement term $\Delta
\varepsilon_{re}=$-0.16 MeV fm$^{-3}$, the difference between the
total energy densities is $\Delta\varepsilon=$0.01 MeV fm$^{-3}$,
i.e., the difference between their energy per nucleon $\Delta
E_B/A=\Delta(\varepsilon/\rho_0-m)=$0.07 MeV, which is small
compared with $E_B/A\approx$ 16 MeV.

The saturation properties of symmetric nuclear matter for
different interactions including the Fermi momenta $k_F$,
saturation densities $\rho_0$, energies per nucleon $E_B/A$,
effective masses $m^*$ and $m^*/m$, incompressibility $K$ and
symmetric energy coefficients $a_{sym}$ are shown in Table
\ref{tab:table2}. We can see that the results given by the
density-dependent interactions DD-ME1 and TW-99 are similar to
those of the nonlinear interactions except for TM2, NL2 and GL-97.
The saturation density for TM2 is lower and the energy per nucleon
for NL2 is larger than the corresponding empirical values, which
can also be seen in Fig. \ref{fig:fig6}. Due to the weakest scalar
potentials shown in the middle panel of Fig. \ref{fig:fig3}, GL-97
gives the largest effective mass $m^*/m$=0.78 and
incompressibility $K$=240 MeV, which are justified from the
empirical nuclear saturation properties \cite{glen97}.

Although properties below the saturation densities for symmetric
nuclear matter are quite similar, the EOS at low densities given
by various interactions for pure neutron matter is quite
different. This is due to the effective interactions used so far
are obtained by fitting the properties of doubly magic nuclei,
which have an isospin close to that of the symmetric nuclear
matter. To get an EOS for neutron matter without any ambiguity, it
is necessary to constrain the effective interactions either by
microscopic many-body calculations for the neutron matter data
\cite{heis00} or the data of nuclei with extreme isospin.

\subsection{Properties of neutron stars}

\begin{figure}[ht]
 \centering
 \includegraphics[scale=0.5]{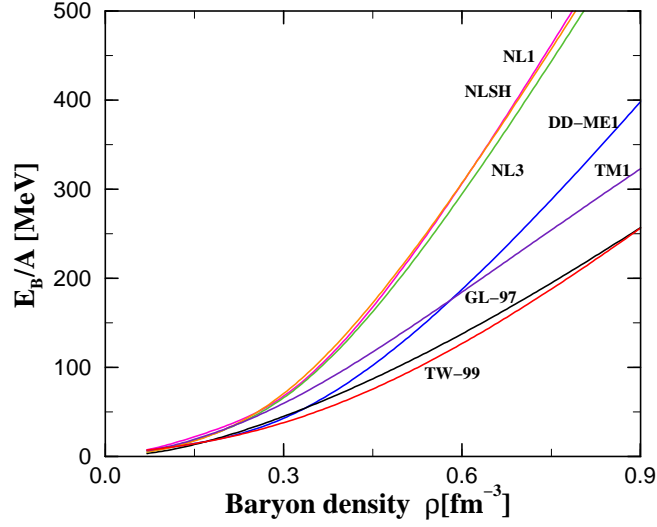}
 \caption{(Color online) Similar as Fig. \ref{fig:fig5}, but for
neutron stars.}
 \label{fig:fig7}
\end{figure}

 The energies per baryon for neutron stars as
functions of baryon density for different interactions are given
in Fig. \ref{fig:fig7}. At low densities, Fig. \ref{fig:fig7} is
very similar to the top panel of Fig. \ref{fig:fig6} as the
neutron stars matter is almost pure neutron matter at low
densities.  The energies per baryon for the nonlinear interactions
NL1, NL3 and NLSH increase quickly with the density compared with
those of TW-99 and GL-97, while the results for DD-ME1 and TM1 lie
in-between. These results are in consistent with the total
potentials for neutron stars in Fig. \ref{fig:fig5}. At density
$\rho\sim 0.3$ fm$^{-3}$, the potentials for TW-99 and DD-ME1 are
the lowest, while that of DD-ME1 crosses GL-97 and TM1 at
$\rho\sim 0.4$ fm$^{-3}$ in Fig. \ref{fig:fig5}. The $E_B/A$ for
TW-99 is  the smallest.

\begin{figure}[ht]
 \centering
 \includegraphics[width=12cm]{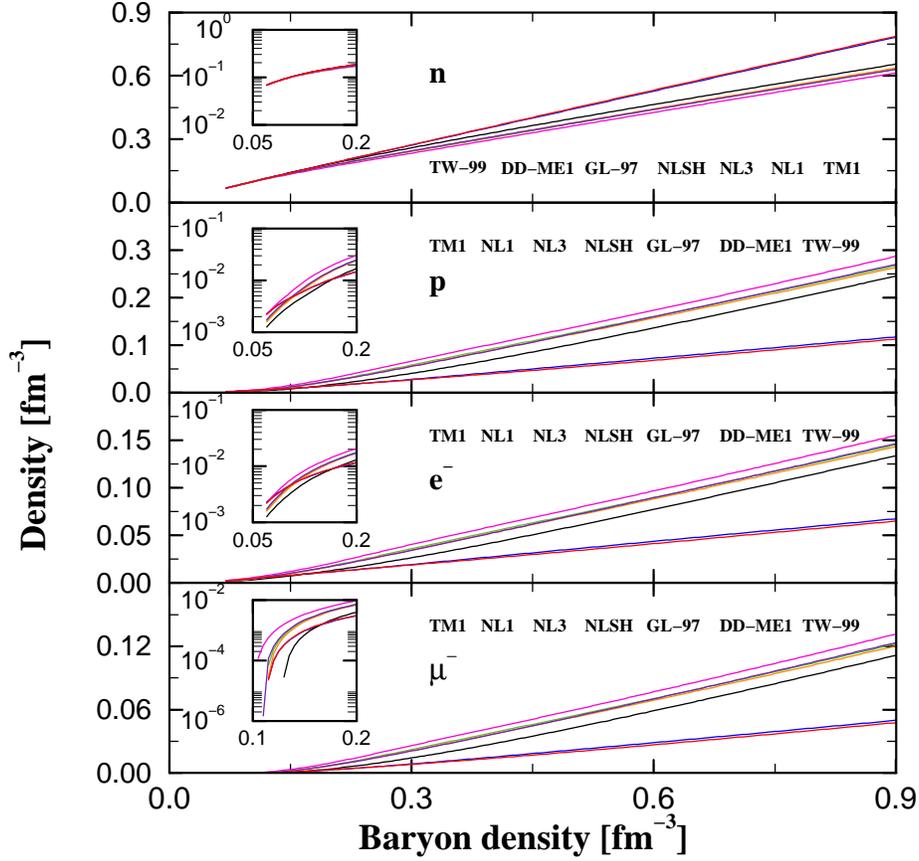}
 \caption{(Color online) Particle densities (from top to bottom are
respectively n, p, $e^-$ and $\mu^-$ ) in neutron stars for
different effective interactions as a function of the baryon
density $\rho$. The corresponding inserts are the same figures in
a logarithmic scale at low density ($0.05<\rho$ fm$^3<0.2$). The
curves are labelled from the top to the bottom at $\rho$=0.45
fm$^{-3}$ orderly from left to right.}
 \label{fig:fig8}
\end{figure}

The evolutions of particle densities with the baryon density in
neutron stars are given in Fig. \ref{fig:fig8},  and the
corresponding figures in logarithm scale for low densities
($0.05<\rho$ fm$^3<0.2$) are given in the sub-figures, where the
solutions are from $\rho$=0.065 fm$^{-3}$ to 0.9 fm$^{-3}$, i.e.
$0.425\leq\rho_B/\rho_0\leq5.88$ with $\rho_0=0.153$ fm$^{-3}$.

At low densities, the charge-neutral neutron star matter is mainly
composed of neutrons. As the density increases, high-momentum
neutrons  will $\beta$-decay into protons and electrons (
$n\leftrightarrow p+e^- +\overline{\nu}_e$ ) until the equilibrium
at which the chemical potentials satisfy $\mu_p=\mu_n-\mu_e$.  As
the neutron density increases, so do the proton and electron
densities. When $\mu_e$ attains the value of the muon mass, the
$\mu^-$ will appear. The equilibrium with respect to the reaction
$e^-\leftrightarrow\mu^- +\overline{\nu}_\mu+\nu_e$ implies that
$\mu_e=\mu_\mu$. The $\mu^-$ thresholds are different for
different effective interactions. All the $\mu^-$ thresholds are
in the range of $\rho=0.11\pm0.01$ fm$^{-3}$ with the minimum and
maximum thresholds given by NL1 and GL-97 respectively.

From Fig. \ref{fig:fig8}, it can also be seen that the
density-dependent effective interactions TW-99 and DD-ME1 give the
largest neutron densities, and accordingly the smallest proton
densities due to the baryon number conservation. This is because
the strengths $g_{\rho B}$ for the density-dependent effective
interaction become weaker with the baryon density, as shown in
Fig. \ref{fig:fig2}. Because of the charge-neutral condition, the
densities of electron and $\mu^-$ for different effective
interactions have the same tendencies as the proton densities.

As hyperons would appear at roughly twice saturation density, it
is necessary to sudy neutron star with hyperons. The details for
the inclusion of hyperons in neutron star in relativistic mean
field theory are given in Refs.\cite{glen97, prak97}. One can
introduce the ratios of the meson- hyperons coupling strengths
coupling constants to those of nucleons as: \beqn
 x_{\sigma h}=\frac{g_{\sigma h}}{g_{\sigma N}},~~~
 x_{\omega h}=\frac{g_{\omega h}}{g_{\omega N}},~~~
 x_{\rho h}  =\frac{g_{\rho h}}{g_{\rho N}},
 \eeqn
where, $g_{\sigma h}, g_{\omega h}$ and $g_{\rho h}$ have the same
density dependencies as $g_{\sigma N}, g_{\omega N}$ and $g_{\rho
N}$ respectively, and the ratios $x_{\sigma h}=x_{\omega
h}=x_{\rho h}=\sqrt{2/3}$ are chosen according to Ref.
\cite{mosz74}.

\begin{figure}[ht]
 \centering
 \includegraphics[width=9cm]{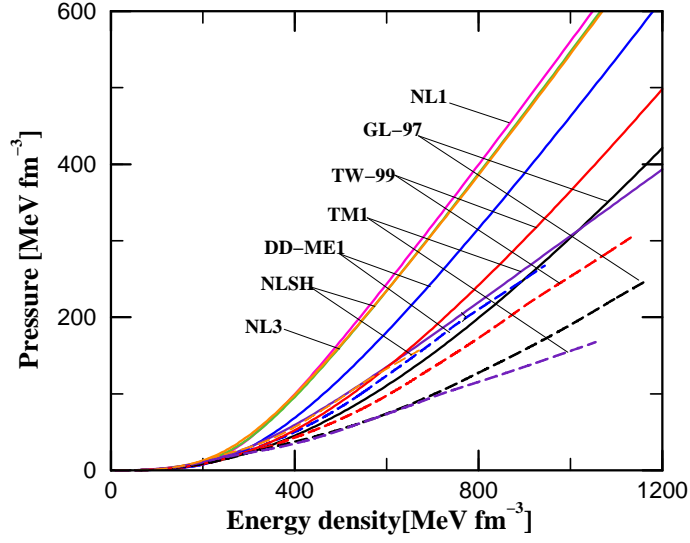}
 \caption{(Color online) EOS of neutron stars for different
effective interactions (as marked in figure). The solid lines
represent these with neutrons and protons only, and the dashed
lines represent the corresponding ones with hyperons included
respectively.}
 \label{fig:fig9}
\end{figure}

The equation of state (EOS) is very important to understand the
structure of neutron star. The stiffer the EOS, the larger the
mass that can be sustained against collapse. There are two
constraints for the realistic EOS. One is a stiff limit by the
causal constraint $\partial p /
\partial \varepsilon \leq 1$, which results in the limit mass of just over
3M$_\odot$. The other corresponds to the soft limit, which
corresponds to the free Fermi gas with neutrons, protons and
leptons in equilibrium and the limit mass is about $0.7M_\odot$
\cite{glen97}. Here the EOS calculated by different effective
interactions are given in Fig. \ref{fig:fig9}. The solid lines
represent these with neutrons and protons only, and the dashed
lines represent their corresponding ones with hyperons included
respectively.  As can be understood from the potentials in
Fig.\ref{fig:fig5},  the nonlinear interactions NL1, NL3 and NLSH
give stiffer EOS than the other interactions,  GL-97  and TM1 give
softer EOS, and density-dependent interactions TW-99 and DD-ME1
lie in-between. Furthermore, the inclusion of hyperons softens the
corresponding EOS considerably, as shown by the corresponding
dashed lines in Fig. \ref{fig:fig9}.  The softest EOS is given by
TM1. After the inclusion of hyperons, the corresponding solutions
for the nonlinear interactions, NL1, NL3 and NLSH, exist only
below density, $\rho=$ 0.42, 0.51, and 0.58 fm$^{-3}$,
respectively. Beyond the corresponding density, the scalar
potential $g_{\sigma B}\sigma$ will increase and make the
effective masses $m^*=m+g_{\sigma B}\sigma$ negative. Therefore in
the following, we do not discuss the properties of neutron stars
with hyperons for NL1, NL3 and NLSH.

\begin{figure}[ht]
 \centering
 \includegraphics[width=14cm]{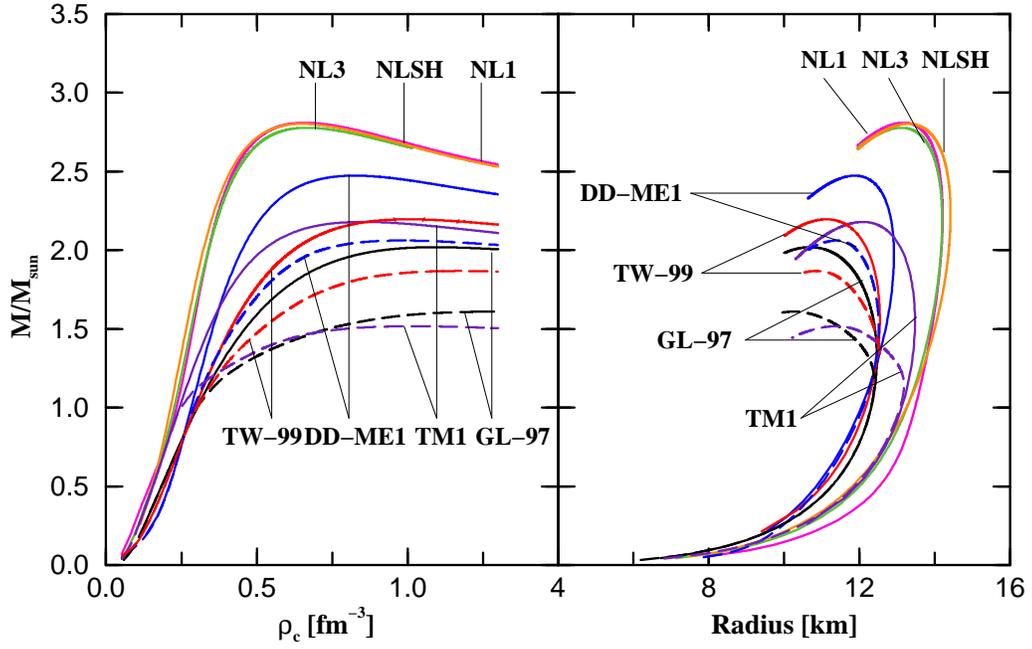}
 \caption{(Color online) The masses versus the central densities (left panel) and
radii (right panel) in neutron stars for different effective
interactions (as marked in figure). The solid lines represent
these with neutrons and protons only, and the dashed lines
represent the corresponding ones with hyperons included.}
 \label{fig:fig10}
\end{figure}

For a static global star, the Oppenheimer-Volkoff-Tolman (OVT)
equation is \cite{oppe39,tolm39}:
  \beqn
    \frac{dp}{dr}&=&-\frac{[p(r)+\varepsilon(r)][M(r)+4\pi r^3p(r)]}{r(r-2M(r))}\\
     &&M(r)=4\pi\int_0^r\varepsilon(r)r^2dr
  \eeqn
The point $R$, at which the pressure vanishes, $p(R)=0$, defines
the radius of the star and $M(R)$ is the gravitational mass. For a
given EOS, the OVT equation has a unique solution which depends on
a single parameter characterizing the conditions of matter at the
center. This can be chosen as the baryon density or energy
density. In Fig. \ref{fig:fig10}, the masses versus the central
densities (left panel) and radii (right panel) of neutron stars
for different interactions are shown.  The solid lines represent
these with neutrons and protons only, and the dashed lines
represent the corresponding ones with hyperons included. There is
critical maximum value for the masses of neutron star, known as
the OV mass limit. Beyond this mass the star is unstable to
gravitational collapse.

The OV mass limits, corresponding radii, central densities, energy
densities and pressures for different effective interactions are
presented in Table \ref{tab:table3}. The second rows for DD-ME1,
TW-99, GL-97 and TM1 represent the quantities with hyperons.
Without hyperons, the OV mass limits calculated from different EOS
are 2.02 $\sim$ 2.81$ M_\odot$, and the corresponding radii are
10.24 $\sim$ 13.27 km. The OV mass limits and corresponding radii
for DD-ME1 and TW-99 are respectively $2.475 M_\odot,11.903$ km
and $2.195 M_\odot,11.209$ km and they give softer EOS and smaller
OV mass limits than those of the effective interactions NL1
($2.809 M_\odot,13.137$ km), NL3 ($2.778 M_\odot,13.081$ km) and
NLSH ($2.803 M_\odot,13.270$ km). The results for TM1 are similar
to TW-99, while GL-97 gives the smallest OV mass limit ($2.018
M_\odot$) and radius (10.779 km).

Due to the softer EOS for neutron star with hyperons, small OV
mass limits have been obtained. The OV mass limits calculated with
hyperons for DD-ME1, TW-99 GL-97 and TM1 are 1.52 $\sim$ 2.06$
M_\odot$, and the corresponding radii are 10.24 $\sim$ 11.38 km.
With hyperons included, the OV mass limits and corresponding radii
for DD-ME1 and TW-99 are respectively $2.061 M_\odot,11.375$ km
and $1.868 M_\odot,10.853$ km. The results for TM1 are similar to
TW-99, and gives radius (11.366 km) and the smallest OV mass limit
($1.517 M_\odot$).


\section{Summary}

We have studied and carefully compared the density dependencies of
various effective interaction strengths in symmetric nuclear
matter, pure neutron matter and neutron stars. The corresponding
influences on potentials and properties of symmetric nuclear
matter, pure neutron matter and neutron stars are presented and
discussed. As the interactions NL2 and TM2 are aimed for light
nuclei, we don't present their results for neutron star. The
properties calculated by the interactions NL1, NL3 and NLSH are
close to each other. The same conclusion can be seen for the
density-dependent interactions TW-99 and DD-ME1.

For the $\sigma$-meson, all the interaction strengths are
density-dependent. While for the $\omega$-meson, TW-99, DD-ME1 and
TM1 are density-dependent. For the $\rho$-meson, the interaction
strengths for TW-99 and DD-ME1 decrease very fast with density
while the others are constant. Even though the interaction
strengths and the potentials from various interactions are
different, the differences of saturation properties for various
effective interactions in symmetric nuclear matter are subtle
except for NL2 and TM2.

Unlike those for symmetric nuclear matter, the properties for pure
neutron matter by various interactions are quite different. As the
effective interactions used so far are obtained by fitting the
properties of doubly magic nuclei, it may be successful for
nuclear matter with an isospin close to that of the symmetric
nuclear matter. To get an EOS for neutron matter without any
ambiguity, it is necessary to constrain the effective interactions
either by microscopic many-body calculations for the neutron
matter data or the data of nuclei with extreme isospin.

For neutron star matter, the density-dependent interactions DD-ME1
and TW-99 favor large neutron populations due to their weak
$\rho$-meson field at high densities. The OV mass limits
calculated from different EOS are 2.02 $\sim$ 2.81$M_\odot$, and
the corresponding radii are 10.78 $\sim$ 13.27 km. The stronger
interaction gives a stiffer EOS and a larger mass limit. TW-99 and
DD-ME1 give softer EOS and smaller OV mass limits than those of
the effective interactions NL1, NL3 and NLSH. The results of TM1
are similar to TW-99. After the inclusion of hyperons, the
corresponding values become 1.52 $\sim$ 2.06 $M_\odot$ and 10.24
$\sim$ 11.38 km.

\begin{acknowledgments}

We would like to express our gratitude to G. C. Hillhouse for his
careful reading of the manuscript. This work is partly supported
by the Major State Basic Research Development Program under
Contract Number G2000077407 and the National Natural Science
Foundation of China under Grant No. 10025522, 10221003, and
10047001.
\end{acknowledgments}


\renewcommand{\baselinestretch}{1} \small \normalsize

 \begin{table}[ht!]
  \centering
  \caption{The Fermi momenta $k_F$ ($\rho_0={k_F^3}/{3\pi^2}$),  vector
densities $\rho_0$, scalar densities $\rho_s$, scalar potentials
$g_{\sigma B}\sigma$, vector potentials $g_{\omega B}\omega_0$,
the energy densities $\varepsilon_\sigma$, $\varepsilon_\omega$,
$\varepsilon_{re}$, $\varepsilon_N^p$, $\varepsilon_N^k$, and
$\varepsilon_N$, the total energy densities $\varepsilon$, nucleon
masses $m$ and energies per nucleon $E_B/A=\varepsilon/\rho-m$ for
different interactions in symmetric nuclear matter at saturation
density (the units are in MeV$\cdot$ fm$^{-3}$ except otherwise
stating).}
   \label{tab:table1}
   \begin{ruledtabular}
   \begin{tabular}{c|c|c|c|c|c|c|c|c|c}
 &\multicolumn{1}{c|}{TW-99}
 &\multicolumn{1}{c|}{DD-ME1} &\multicolumn{1}{c|}{GL-97}
 &\multicolumn{1}{c|}{NL1} &\multicolumn{1}{c|}{NL2}
 &\multicolumn{1}{c|}{NL3} &\multicolumn{1}{c|}{NLSH}
 &\multicolumn{1}{c|}{TM1}&\multicolumn{1}{c}{TM2}\\
 \hline
  $k_F$(fm$^{-1}$)&1.313&1.310&1.313&1.310&1.292&1.300&1.293&1.291&1.251\\
  $\rho_0$(fm$^{-3}$)&0.153&0.152&0.153&0.152&0.146&0.148&0.146&0.145&0.132\\
  $\rho_s$(fm$^{-3}$)&0.143&0.143&0.148&0.142&0.139&0.140&0.138&0.138&0.125\\
 \hline
  $g_{\sigma B}\sigma$(MeV)
  &-417.9&-396.2&-206.4&-400.7&-309.7&-380.3&-378.24&-342.94&-402.06\\
  $-\frac{1}{2}g_{\sigma B}\sigma\rho_s$
  &29.871&28.279&15.274&28.450&21.524&26.621&26.099&23.663&25.129\\
  $\varepsilon_\sigma$&29.871&28.279&13.830&28.187&20.538&26.088&25.544&22.387&24.027\\
 \hline
  $g_{\omega B}\omega_0$(MeV)
  &338.7&316.7&145.57&325.4&242.9&308.0&306.6&274.5&331.8\\
  $-\frac{1}{2}g_{\omega B}\omega_0\rho_0$
  &-25.910&-24.069&-11.140&-24.689&-17.688&-22.832&-22.378&-19.901&-21.897\\
  $\varepsilon_\omega$&-25.910&-24.069&-11.140&-24.689&-17.688&-22.832&-22.378&-19.408&-21.221\\
 \hline
  $\varepsilon_{re}$&-0.161&-0.364&0&0&0&0&0&0&0\\
 \hline
  $\varepsilon_N^p$&51.981&48.502&22.281&49.378&35.375&45.664&44.757&39.856&43.893\\
  $\varepsilon_N^k$&85.401&87.912&116.199&87.207&96.244&87.643&86.786&91.016&75.261\\
  $\varepsilon_N$&137.382&136.414&138.470&136.585&131.619&133.307&131.543&130.872&119.155\\
 \hline
  $\varepsilon$&141.182&140.260&141.169&140.083&134.470&136.563&134.708&133.649&120.678\\
  $m$(MeV)&939&939&939&938&938&939&939&938&938\\
 \hline
  $E_B/A$(MeV)&-16.25&-16.23&-16.32&-16.42&-17.02&-16.25&-16.35&-16.265&-16.16\\
  \end{tabular}
  \end{ruledtabular}
 \end{table}

\begin{table}[ht]
 \centering
 \caption{Nuclear matter saturation properties for
different effective interactions, including the Fermi momenta
$k_F$, saturation densities $\rho_0$, energies per nucleon
$E_B/A$, effective masses m$^*$ and m$^*$/m, incompressibility K
and symmetric energy coefficients $a_{sym}$. } \label{tab:table2}
\begin{ruledtabular}
\begin{tabular}{c|c|c|c|c|c|c|c}
&\multicolumn{1}{c|}{$k_F$(fm$^{-1}$)}&\multicolumn{1}{c|}{$\rho_0$(fm$^{-3}$)}
&\multicolumn{1}{c|}{$E_B/A$(MeV)}&\multicolumn{1}{c|}{m$^*$(MeV)}&\multicolumn{1}{c|}{m$^*$/m}
&\multicolumn{1}{c|}{$K$(MeV)}&\multicolumn{1}{c}{$a_{sym}$(MeV)}\\
\hline
DD-ME1&1.310&0.152&-16.23&542.7899&0.578&244.50&33.06\\
TW-99&1.313&0.153&-16.25&521.0724&0.555&240.00&32.77\\
NL1&1.310&0.152&-16.42&537.2949&0.573&211.15&43.47\\
NL2&1.292&0.146&-17.02&628.3138&0.670&399.17&45.12\\
NL3&1.300&0.148&-16.25&558.6835&0.595&271.73&37.42\\
NLSH&1.293&0.146&-16.35&560.7559&0.598&355.34&36.12\\
TM1&1.291&0.145&-16.265&595.0626&0.634&281.17&36.89\\
TM2&1.251&0.132&-16.16&535.9376&0.571&343.83&35.98\\
GL-97&1.313&0.153&-16.32&732.6145&0.780&240.00&32.5\\
  \end{tabular}
  \end{ruledtabular}
\end{table}


\setlength{\tabcolsep}{2pt}
\begin{table}[ht]
\centering \caption{ The central densities, energy densities, and
pressures, OV mass limits and corresponding radii for neutron
stars for different effective interactions. For interactions
DD-ME1, TW-99, TM1 and GL-97, the same quantities for neutron
stars with hypersons are given in the following lines
respectively.} \label{tab:table3}
\begin{ruledtabular}
\begin{tabular}{c|c|c|c|c|c}
 &\multicolumn{1}{c|}{Central density}
 &\multicolumn{1}{c|}{Central energy density}
 &\multicolumn{1}{c|}{central pressure}
 &\multicolumn{1}{c|}{OV mass limit}
 &\multicolumn{1}{c}{Radius}\\
 &(fm$^{-3}$)&($\times10^{15}$g/cm$^3$)
 &($\times10^{35}$dyne/cm$^2$)
 &($\times M_\odot$)&(km)\\
\hline
DD-ME1&0.815&1.852&7.886&2.475&11.903\\
      &0.980&2.097&5.780&2.061&11.375\\
TW-99 &0.970&2.126&7.902&2.195&11.209\\
      &1.176&2.422&6.318&1.868&10.853\\
TM1   &0.852&1.881&5.293&2.180&12.054\\
      &1.016&2.083&3.071&1.517&11.366\\
GL-97 &1.045&2.347&7.905&2.018&10.779\\
      &1.299&2.792&6.188&1.610&10.242\\
NL1   &0.658&1.529&7.140&2.809&13.137\\
NL3   &0.667&1.544&7.055&2.778&13.081\\
NLSH  &0.649&1.497&6.682&2.803&13.270\\
 \end{tabular}
 \end{ruledtabular}
\end{table}

\newpage


\end{document}